\documentclass[10pt,conference]{IEEEtran}
\usepackage{cite}
\usepackage{amsmath,amssymb,amsfonts}
\usepackage{algorithmic}
\usepackage{graphicx}
\usepackage{textcomp}
\usepackage{xcolor}
\usepackage[hyphens]{url}
\usepackage{fancyhdr}
\usepackage{hyperref}
\usepackage[ruled,vlined]{algorithm2e}
\usepackage{multirow}
\usepackage{subfig}

\begin{document}

\title{New Covert and Side Channels Based on Retirement}

\author{\IEEEauthorblockN{Ke~Xu\IEEEauthorrefmark{2},
Ming~Tang\IEEEauthorrefmark{1}\IEEEauthorrefmark{2},
Quancheng~Wang\IEEEauthorrefmark{2},
Han~Wang\IEEEauthorrefmark{2}}
\IEEEauthorblockA{
\IEEEauthorrefmark{2} Key Laboratory of Aerospace Information Security and Trusted Computing,
Ministry of Education,\\
School of Cyber Science and Engineering, Wuhan University, Wuhan, 430072, China
}
\IEEEauthorblockA{\{kexuwhu, m.tang, wangquancheng, han.wang\}@whu.edu.cn}
}

\maketitle
\thispagestyle{plain} 
\begin{abstract}

  Intel processors utilize the retirement to orderly retire the µops that have been executed out of order. To enhance retirement utilization, the retirement is dynamically shared between two logical cores on the same physical core. However, this shared retirement mechanism creates a potential vulnerability wherein an attacker can exploit the competition for retirement to infer the data of a victim on another logical core on the same physical core. Based on this leakage, we propose two new covert channels: the Different Instructions (DI) covert channel using different instructions for information transmission, and the Same Instructions (SI) covert channel using the same instructions to transmit information. The DI covert channel can achieve $98.5\%$ accuracy with a bandwidth of $1450$ Kbps, while the SI covert channel can achieve $94.85\%$ accuracy with a bandwidth of $483.33$ Kbps. Furthermore, this paper explores additional applications of retirement: Firstly, retirement is applied to Spectre attacks, resulting in a new variant of Spectre v1, which can achieve $94.17\%$ accuracy with a bandwidth of $29$ Kbps; Secondly, retirement is leveraged to infer the programs being executed by the victim, which can infer $10$ integer benchmarks of SPEC with $89.28\%$ accuracy. Finally, we discuss possible protection against new covert channels and other applications of retirement.

\end{abstract}

\section{Introduction}
\label{section1}
Modern processors incorporate various optimization strategies aimed at enhancing overall performance, 
while sharing microarchitectural resources among different logical cores to improve resource utilization. 
However, the optimization strategies cannot optimize all instructions, and the shared strategies can 
introduce contention stalling, resulting in observable execution time differences for the same instruction. Exploiting this principle, existing attackers have 
devised covert channels capable of bypassing privilege boundaries to transmit sensitive information. 
In such covert channel attacks, the sender leverages the modification of microarchitectural resource 
states to encode the private information, while the receiver decodes the information by analyzing the 
execution time of specific instructions. Furthermore, side channel 
attacks~\cite{6bernstein2005cache,9yarom2014flush+,10gruss2016flush+,14xiao2023exploiting,15guo2022adversarial,16gras2018translation,20yarom2017cachebleed,26aldaya2019port,27gast2022squip}, 
Spectre attacks~\cite{31kocher2020spectre}, and Meltdown attacks~\cite{30lipp2020meltdown} reveal 
how covert channels can further cause security threats. In this paper, we propose a novel covert channel 
based on the retirement in Intel processors.

The Intel processor's frontend decodes instructions into 
µops and sends them to the backend for execution. The 
backend executes the µops out of order to reduce stalling. 
To ensure correct program execution, the backend utilizes 
the Reorder Buffer (ROB) to reorder completed µops and uses 
retirement to retire µops in order. If the first µop in the 
ROB has not been completed, retirement will stall until it 
is completed. Retirement will also stall when the ROB is 
empty. Furthermore, retirement is alternately used between 
two logical cores of the same physical core. If one logical 
core does not require retirement, the other logical core 
can continue using retirement. An attacker can exploit this 
behavior to infer the victim's retirement usage on another 
logical core of the same physical core by observing its own 
retirement usage, thereby obtaining the victim's secret.

Based on the above principle, we propose two novel covert channels: the Different Instructions (DI) covert channel and the Same Instructions (SI) covert channel. In both covert channels, the sender and receiver are located on separate logical cores within the same physical core. The sender in each covert channel induces different retirement stalling based on the transmitted data, while the receiver tests the availability of its own retirement by executing instructions that stall at retirement, and then decodes the transmitted data. However, the DI covert channel generates different stalling by executing instructions with varying execution times, and the SI covert channel generates different stalling by executing the same instructions with the required data stored in different locations. The DI covert channel can achieve $98.5\%$ accuracy with a bandwidth of $1450$ Kbps, and the SI covert channel can achieve $94.85\%$ accuracy with a bandwidth of $483.33$ Kbps.

Various protection strategies have been proposed to mitigate the risks posed by existing covert channels. These strategies include partitioning~\cite{3taram2022secsmt,4townley2019smt}, randomization~\cite{32liu2014random,33deng2019secure}, refreshing microarchitectural resources~\cite{34zhang2013duppel,35godfrey2013server}, and detection mechanisms utilizing the Hardware Performance Counters (HPC)~\cite{36kulah2019spydetector,37mushtaq2019sherlock}. However, these protection strategies are designed for specific microarchitectural resources and cannot protect against covert channels that use new microarchitectural resources to transmit information, so the new covert channels proposed in this paper can bypass the existing protection strategies.

Furthermore, we explore further applications of the retirement-based leakage. Leveraging the observation that the branch misprediction penalty exhibits significant variation [2], we applies retirement to the Spectre attack and proposes a new variant of Spectre v1. This variant can achieve an accuracy of $93.16\%$ with a bandwidth of $5.8$ Kbps. In addition, we introduce a novel attack methodology that utilizes the Convolutional Neural Networks (CNN) to infer different programs based on the differences in retirement stalling patterns caused by different program instructions. The proposed attack achieves a success rate of $89.28\%$ in inferring the running programs, using a set of $10$ different integer benchmarks from the SPEC suite as examples of running programs (some benchmarks are grouped together).

The contributions of this paper are as follows:
\begin{itemize}
\item We discover a security vulnerability in the retirement and verify the leakage principle of this vulnerability. 
\item We create two novel covert channels using retirement, and evaluate their performance.
\item We propose a novel variant of the Spectre v1 attack that leverages the retirement and the fallback behavior in misprediction.
\item We apply the retirement to the side channel attacks and develop a method to distinguish between different running programs based on retirement behavior.
\end{itemize}

In the subsequent sections of this paper, Section~\ref{section2} presents the necessary microarchitectural background and the related work. Section~\ref{section3} focuses on validating the leakage principle of the retirement. In Section~\ref{section4}, we build the DI covert channel and the SI covert channel based on retirement, and evaluate the performance of these covert channels. In Section~\ref{section5}, we delve into the exploitation of retirement leakage, proposing a new variant of the Spectre v1 attack and an attack method that leverages retirement to infer different programs. Mitigation strategies are discussed in Section~\ref{section6}, and finally, we conclude the paper in Section~\ref{section7}.

\section{Background and related work}
\label{section2}
\subsection{Microarchitectural structure}

\begin{figure}[!t]
  \centering
  \includegraphics[width=3in]{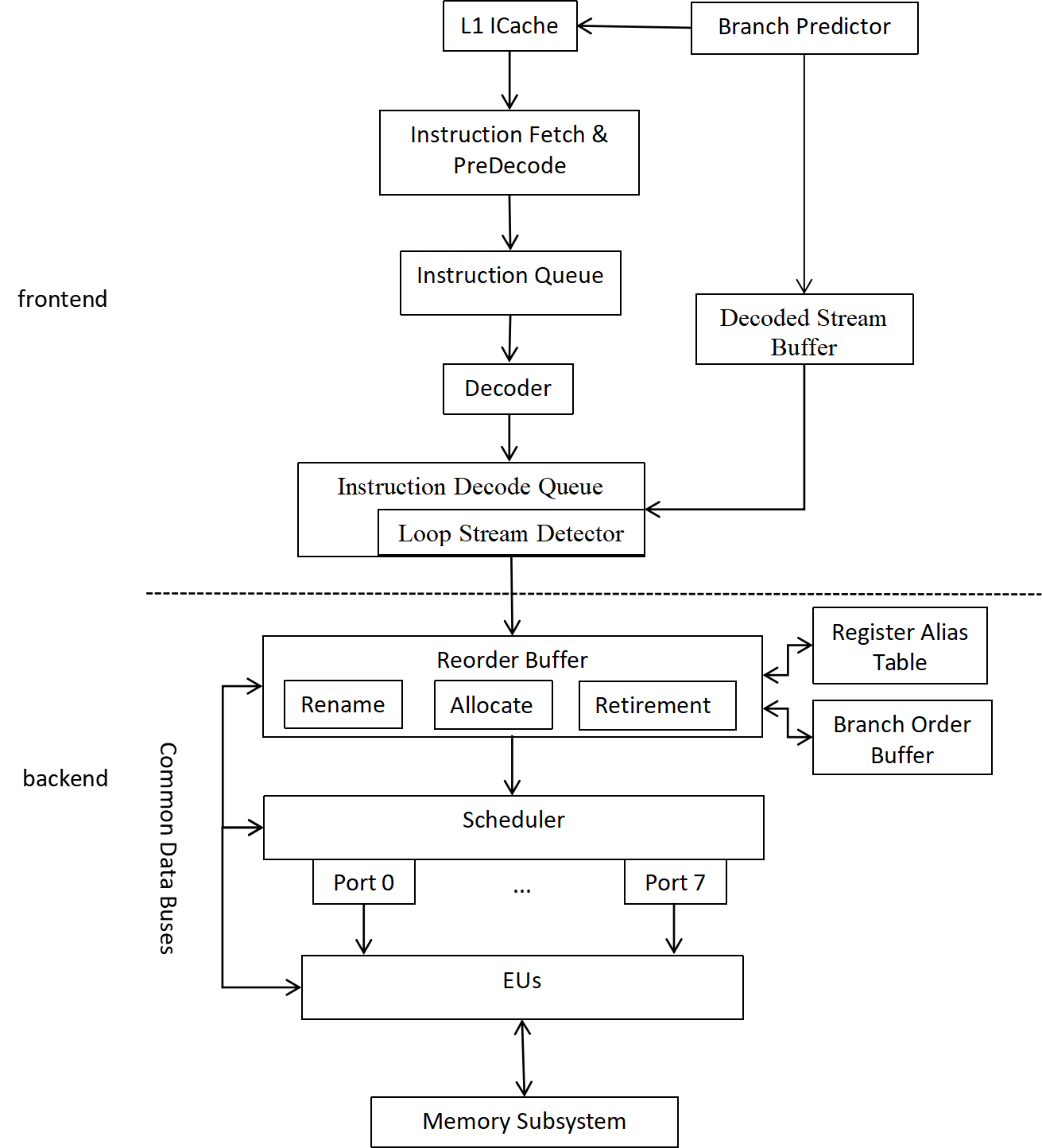}
  \caption{Pipeline structure of Skylake.}
  \label{f1}
\end{figure}

In this subsection, we present the pipeline structure of Intel's existing microarchitecture, taking Skylake as an illustrative example (refer to Figure~\ref{f1}). It's important to note that the overall pipeline structure of Intel's latest microarchitecture remains unchanged compared to Skylake.

The pipeline of Skylake can be categorized into three main components: the frontend, the backend, and the memory subsystem. The frontend is responsible for the crucial task of decoding instructions into µops. The backend component takes charge of executing these µops. Lastly, the memory subsystem plays a crucial role in handling data storage. It ensures efficient and reliable data storage within the microarchitecture. 

The frontend contains three different decoding paths, including the Loop Stream Detector (LSD), which is primarily responsible for decoding loops, the Decoded Stream Buffer (DSB), which handles the decoding of instructions that had their corresponding µops stored within it, and the Micro-Instruction Translation Engine, which decodes instructions whose µops are not stored in the DSB. During the decoding process, the processor determines the appropriate decoding path based on the instructions being executed and the contents of the DSB. The decoded µops are then stored in the Instruction Decode Queue (IDQ), which serves as a buffer for the µops before they are dispatched to the backend for execution. However, it's important to note that LSD has been disabled in Skylake and subsequent microarchitectures.

After the IDQ sends µops to the ROB, the Register Alias Table (RAT) comes into play. The RAT is responsible for renaming registers, allocating resources for stores and loads, and determining all possible scheduler ports. Once the RAT has performed the necessary operations, the allocated µops are forwarded to the scheduler. At this stage, the scheduler assigns execution ports to the µops and the µops are then executed, with their memory data accessed from the memory subsystem. The results of the executed µops are subsequently sent back to the ROB and retirement retires the µops in the ROB in order. Therefore, when the first µop in the ROB is not completed, retirement will stall until the first µop is completed.

\subsection{Microarchitecture-based covert and side channels}

Based on the optimization policies and sharing policies implemented in modern processors, various covert and side channels have been proposed in previous research. These channels exploit specific microarchitectural resources to facilitate unauthorized information transfers and expose sensitive data.

In the memory subsystem, covert and side channels have been discovered by various resources. These include Cache~\cite{5aciiccmez2010new,6bernstein2005cache,8guanciale2016cache,9yarom2014flush+,10gruss2016flush+}, Cache lines' LRU state~\cite{11xiong2020leaking} and consistency state~\cite{12yao2018coherence}, Cache Prefetcher~\cite{13cronin2019fetching,14xiao2023exploiting,15guo2022adversarial}, Cache Bank~\cite{20yarom2017cachebleed}, Translation Lookaside Buffer~\cite{16gras2018translation}, Memory Controller~\cite{21semal2020leaky}, Memory Bus~\cite{22wu2014whispers}, Ring~\cite{42paccagnella2021lord}, Fetch Bandwidth~\cite{3taram2022secsmt}, Store to Load~\cite{28islam2019spoiler}, and Line Fill Buffer~\cite{29wang2023bandwidthbreach}.

In the frontend, covert and side channels exploit resources such as Branch Predictor~\cite{17evtyushkin2015covert,18evtyushkin2016jump}, DSB~\cite{23ren2021see}, LSD~\cite{24deng2022leaky}, Decoder~\cite{3taram2022secsmt}, and Frontend Bus~\cite{25xu2022reverse}.

In the backend, microarchitectural resources prone to exploitation include Execution Port~\cite{26aldaya2019port,7bhattacharyya2019smotherspectre}, Reservation Station and Physical Register File~\cite{3taram2022secsmt}, Scheduler Queue~\cite{27gast2022squip}, and AVX unit~\cite{19schwarz2019netspectre}.

The proposed covert channels have been utilized for various unauthorized transfers, such as cross-process transfers~\cite{25xu2022reverse}, cross-VM transfers~\cite{27gast2022squip}, cross-user transfers on the cloud~\cite{22wu2014whispers}, cross-SGX transfers~\cite{24deng2022leaky}, and cross-physical core transfers~\cite{12yao2018coherence}. On the other hand, side channels have been leveraged to recover keys used in common encryption algorithms like AES~\cite{6bernstein2005cache}, RSA~\cite{27gast2022squip}, and EdDSA~\cite{29wang2023bandwidthbreach}.

\subsection{Transient attacks}
Transient attacks exploit the transient execution to bypass security policies, such as boundary detection, to access confidential information, and then transmit the obtained secret back to normal execution through covert channels. Typically, transient attacks can be divided into three steps: (1) Trigger transient execution by training branch predictors or other methods. (2) Access the secret in the transient execution and encode it into the microarchitectural resources. (3) Decode the secret from the microarchitectural resources in the normal execution.

Existing transient attacks can be categorized into several types based on the methods used to trigger transient execution. These include Spectre v1 attacks triggered by the Pattern History Table~\cite{31kocher2020spectre}, Spectre v2 attacks triggered by the Branch Target Buffer~\cite{38chen2019sgxpectre}, Meltdown attacks triggered by fault-based exploits~\cite{30lipp2020meltdown}, Spectre v4 attacks triggered by Store-to-Load operations~\cite{39horn2018speculative}, and Spectre v5 attacks triggered by the Return Stack Buffer~\cite{40koruyeh2018spectre}.

Various microarchitectural resources are utilized by transient attacks to encode secrets. These resources include Cache~\cite{31kocher2020spectre}, Cache Prefetcher~\cite{15guo2022adversarial}, Line Fill Buffer~\cite{29wang2023bandwidthbreach}, DSB~\cite{23ren2021see}, LSD~\cite{24deng2022leaky}, Frontend Bus~\cite{25xu2022reverse}, Execution Port~\cite{7bhattacharyya2019smotherspectre}, etc.

\section{Characterizing retirement}
\label{section3}

In this section, we analyze the behavior of retirement on the Cometlake microarchitecture and validate the leakage caused by retirement.

\subsection{Usage of retirement}
\label{section3.1}
\subsubsection{Bandwidth of retirement}
\label{section3.1.1}
Prior to conducting the verification experiments, we first try to create an instruction segment stalling at the retirement and test the bandwidth of retirement with it. To accomplish this, we employ nop instruction, which does not necessitate any execution resources and only need to be decoded by the frontend and retired by the retirement.

\begin{figure}[!t]
  \centering
  \includegraphics[width=1.8in]{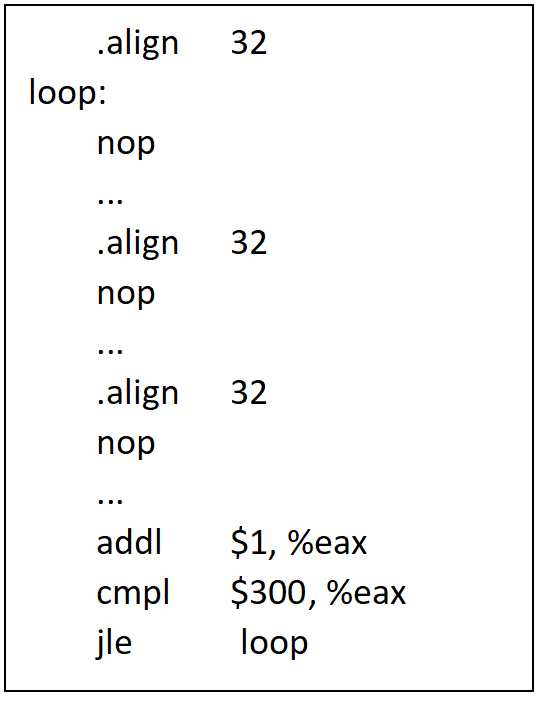}
  \caption{Structure of nop-loop.}
  \label{f2}
\end{figure}

According to~\cite{1coorporation2016intel}, DSB is the fastest decoding method, provided that LSD is disabled. At most one DSB line is decoded per cycle, and each DSB line can store up to 6 µops. Additionally, each $32$-byte instruction corresponds to a maximum of $3$ DSB lines. Therefore, we have designed a specific loop (called nop-loop) and depicted in Figure~\ref{f2}. This loop comprises three $32$-byte instruction blocks, with each block corresponding to three DSB lines, and can be decoded into $18$ µops. This loop ensures the continuous flow of six µops from the frontend to the backend every cycle, thereby eliminating any potential stalling in the frontend.

\begin{figure}[!t]
  \centering
  \includegraphics[width=3in]{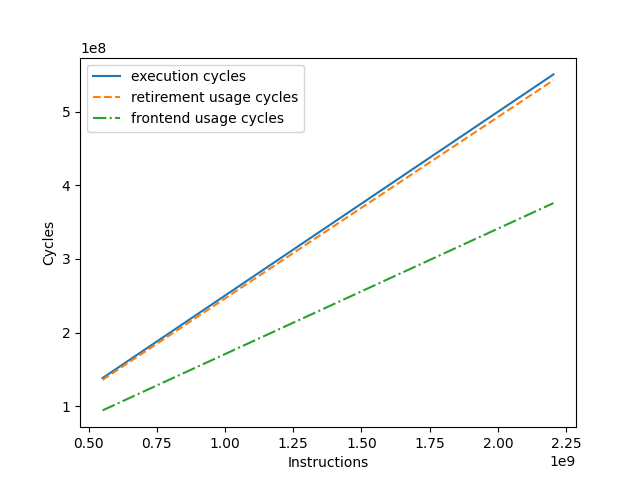}
  \caption{Execution cycles, frontend usage cycles and retirement usage cycles with number of instructions.}
  \label{f3}
\end{figure}

We conducted experiments to test the execution cycles, retirement usage cycles, and frontend usage cycles of the nop-loop by gradually increasing the number of iterations of the loop (i.e., the number of instructions), and the results are presented in Figure~\ref{f3}. It can be observed that the retirement usage cycles are basically the same as the execution cycles, and the frontend usage cycles are significantly smaller than the execution cycles, indicating that the nop-loop takes retirement as the bottleneck, and the execution time is related to the retirement efficiency. The ratio of retirement usage cycles to the number of instructions is approximately 0.25, i.e., the maximum retirement bandwidth is 4 µops per cycle.

\subsubsection{Different instructions lead to different usage of retirement}
\label{section3.1.2}
Retirement retires instructions in order, and retirement will be stalled when the first instruction in the ROB is not completed. Therefore, instructions with different execution cycles will theoretically cause different stalling for retirement. In this subsection, we verify whether instructions with varying execution times result in different patterns of retirement stalling.

\begin{figure}[!t]
  \centering
  \includegraphics[width=2in]{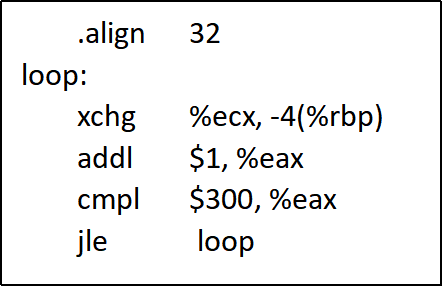}
  \caption{Structure of xchg-loop.}
  \label{f4}
\end{figure}

\begin{figure}[!t]
  \centering
  \includegraphics[width=3in]{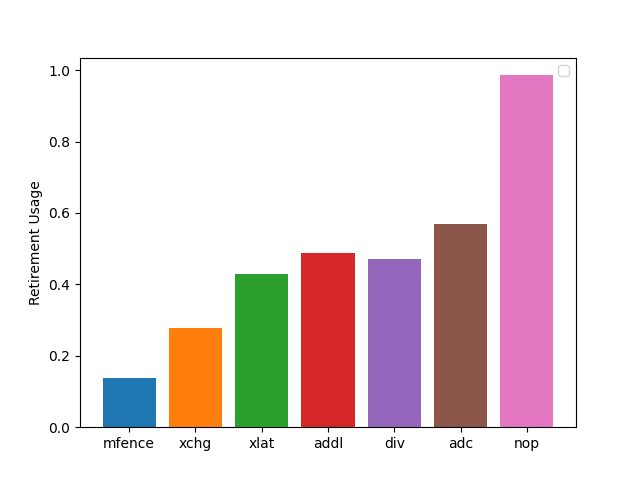}
  \caption{Retirement usage for different instructions.}
  \label{f5}
\end{figure}

We conducted verification experiments using a specific loop structure (called xchg-loop), as illustrated in Figure~\ref{f4}. Each iteration of the xchg-loop generates three µops that can be retired in a single cycle. The xchg of the current iteration and the xchg of the next iteration are interdependent, meaning that the xchg of the next iteration can only begin execution after the completion of the xchg of the current iteration. As a result, each iteration of xchg-loop causes the retirement to generate a stalling time of the execution time of xchg minus one cycle. To investigate the impact of different instructions on retirement stalling, we replaced the xchg instruction with various instructions and measured the retirement usage, which is the ratio of retirement usage cycles to the total cycles of loop execution. In the test experiment, we executed the loops 100,000 times, and the results are depicted in Figure~\ref{f5}.

The experimental results demonstrate that replacing the xchg instruction with different instructions leads to varying levels of retirement usage in the processor. The mfence instruction, which has the longest execution time, exhibits the highest retirement stalling, while the nop instruction, which does not require any execution, shows minimal retirement stalling, almost negligible.

\subsubsection{Same instruction leads to different usage of retirement}
\label{section3.1.3}
The execution time of a particular instruction can vary depending on the location of the data it operates on. This discrepancy in execution time may lead to different stalling in the retirement. Therefore, we conducted experiments to investigate whether the same instruction, when executed with a Cache hit and a Cache miss, would result in different retirement stalling. 

We conduct experiments using the loop structure depicted in Figure~\ref{f4}. We replace the xchg instruction with two different sequences: (1) a flush instruction followed by an instruction that reads data located at the flushed address, and (2) a flush instruction followed by an instruction that reads data from a different address. Each loop iterates $100,000$ times. The experimental results demonstrate that (1) exhibits a retirement usage of $0.0235$, while (2) has a retirement usage of $0.0858$. These findings indicate  that the same instruction can lead to varying retirement stalling depending on the location of the accessed data.

\subsection{Retirement leakage verification}
\label{section3.2}
\subsubsection{Validation of the sharing method of the retirement}
\label{section3.2.1}
\begin{figure}[!t]
  \centering
  \subfloat[Sender uses the retirement.]{\includegraphics[width=3in]{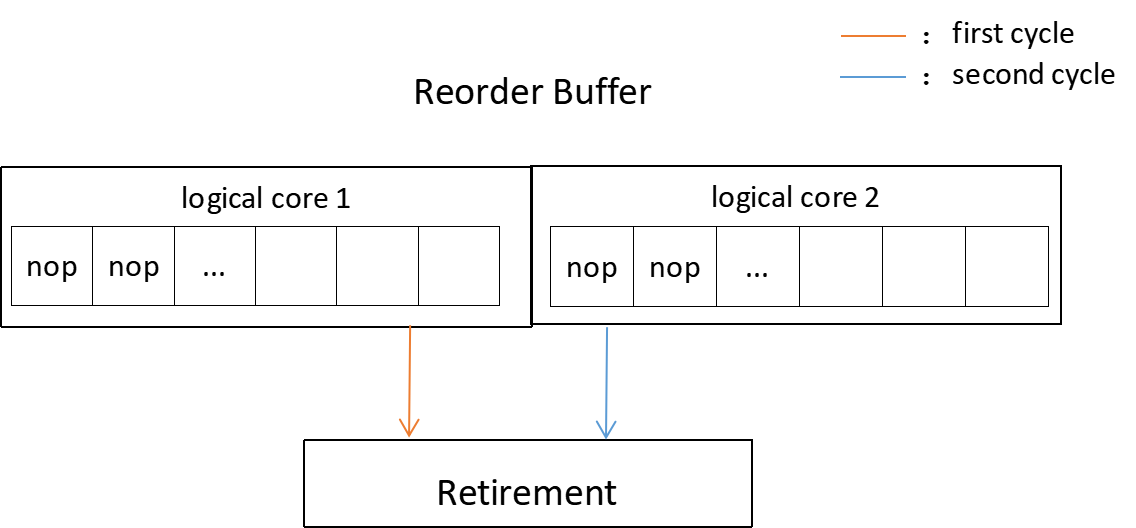}
  \label{fig6a}}
  \hfil
  \subfloat[Sender does not use the retirement.]{\includegraphics[width=3in]{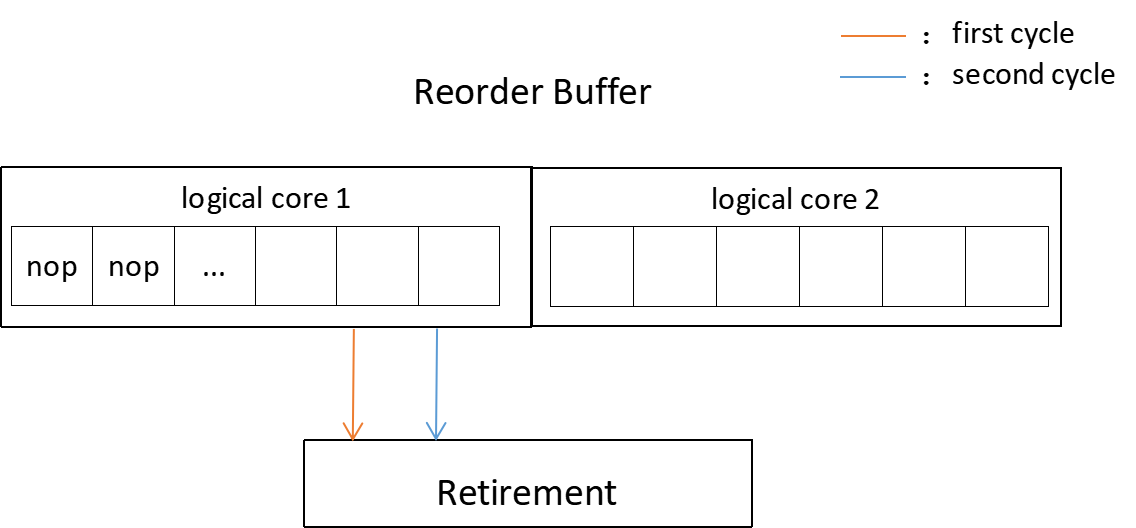}
  \label{fig6b}}
  \caption{Effect of sender on receiver retirement usage.}
  \label{fig6}
\end{figure}

According to the official Intel document~\cite{1coorporation2016intel}, the ROB is partitioned between two logical cores of the same physical core, and retirement is used alternately between them, i.e., dynamically shared. We divide the usage time of the retirement into the first cycle and the second cycle. When both logical cores have instructions ready for retirement, they take turns using the retirement as illustrated in Figure~\ref{fig6a}. On the other hand, if one of the logical cores does not require retirement, the other logical core can utilize both cycles of retirement, as depicted in Figure~\ref{fig6b}.

We treat two logical cores of the same physical core as sender and receiver, and verify whether the retirement is dynamically shared. We run xchg-loop and nop-loop continuously on the sender respectively, so that the sender generates different stalling in the retirement. Simultaneously, on the receiver, we executed nop-loop with $100,000$ iterations to measure its execution cycles, retirement usage cycles, and retirement stalling cycles.

\begin{figure}[!t]
  \centering
  \includegraphics[width=3in]{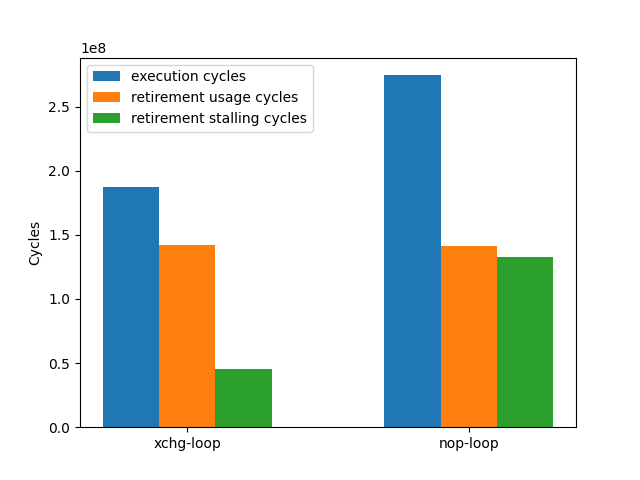}
  \caption{Effect of different loops executed on the sender.}
  \label{f7}
\end{figure}

The results depicted in Figure~\ref{f7} indicate that when the sender executes loops with different retirement stalling times, the receiver's retirement exhibits consistent usage cycles. However, the retirement stalling cycles vary, resulting in different execution times for the nop-loop on the receiver. This observation confirms that retirement is dynamically shared between the two logical cores of the same physical core.

\subsubsection{Execution time differences}
\label{section3.2.2}
After verifying that the retirement is dynamically shared, we proceed to measure the execution time difference that retirement can introduce to the receiver. This measurement is crucial for the subsequent creation of the covert channels.

(1) Differences in execution time due to different instructions

We first test the timing difference introduced by the sender running different instructions on the receiver. The sender and receiver are synchronized using timestamps. Every $10,000$ cycles, the sender executes the xchg-loop with $350$ iterations or the nop-loop with $200$ iterations, while the execution time of the nop-loop with $100$ iterations on the receiver is measured. This configuration ensures that the sender starts executing before the receiver and concludes after the receiver.

\begin{figure}[!t]
  \centering
  \includegraphics[width=3in]{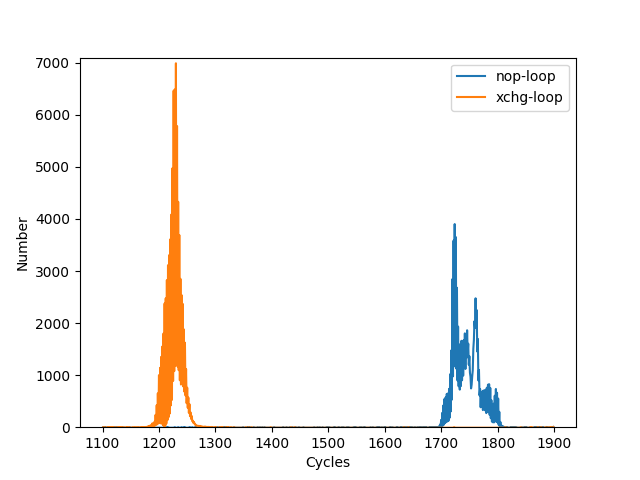}
  \caption{Execution time of receiver's nop-loop when the sender executes different loops.}
  \label{f8}
\end{figure}

Figure~\ref{f8} presents the results, demonstrating that the execution time of the receiver's nop-loop exhibits significant variations when the sender executes different loops with varying retirement stalling.

(2) Differences in execution time due to same instructions

Next, we measure the timing differences induced by the sender running the same instructions with data stored in different locations on the receiver. The sender and receiver are synchronized using timestamps. Every $10,000$ cycles, the sender executes loops with data in the Cache or loops with data not in the Cache. Meanwhile, the execution time of the nop-loop on the receiver is measured. The structure of the loop executed by the sender is depicted in Figure~\ref{f4}, but replacing the xchg instruction with the mov instruction that reads the memory data. The nop-loop on the receiver iterates $100$ times. 

\begin{figure}[!t]
  \centering
  \includegraphics[width=3in]{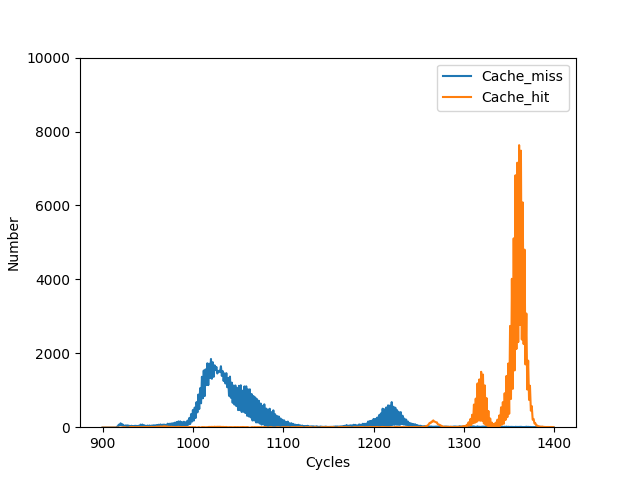}
  \caption{Execution time of receiver's nop-loop when the sender executes the same loop with required data stored in different locations.}
  \label{f9}
\end{figure}

According to our tests, when the number of loop iterations is $60$ or more, i.e., when the read of data involves $60$ Cache lines, the execution time of the receiver's nop-loop exhibits a notable difference. The execution time of the receiver's nop-loop is shown in Figure~\ref{f9} when the sender involves the reading of data of $60$ Cache lines. The results indicate that the nop-loop on the receiver produces different execution times depending on whether the data required for the same loop executed by the sender is in the Cache line or not.

\section{Retirement covert channels}
\label{section4}
\subsection{Threat model}
\label{section4.1}
In our attack scenario, we assume that the sender and the receiver in the attack are two user-level processes running on a processor with hyperthreading enabled. The sender is a Trojan code designed to extract private information from the user and transmit it covertly to the receiver, who is the attacker. Both processes are assigned to different logical cores of a specific physical core using techniques like sched\_setaffinity~\cite{41Chromium2018Speculative}. To ensure successful communication between the sender and receiver, they must establish a unified transmission protocol, including synchronization, allocation, encoding and decoding methods, and error correction, etc.

Besides, a lot of previous studies have been carried out in shared memory pages and thus limited in attack scenarios~\cite{9yarom2014flush+,12yao2018coherence}, in contrast to our study.

Note that the sender is unable to transmit secrets directly to the receiver due to the system security auditors. Hence, the sender needs to employ covert methods to disclose private information.

\subsection{Building covert channels}
\label{section4.2}
\subsubsection{Synchronization}
\label{section4.2.1}
Prior to transmitting information through covert channels, it is crucial to synchronize the sender and receiver to ensure accurate communication. Similar to the previous work~\cite{9yarom2014flush+,10gruss2016flush+,11xiong2020leaking,15guo2022adversarial,25xu2022reverse}, we synchronize the sender and receiver using timestamps.

The sender and receiver utilize the rdtsc instruction to obtain the current timestamp ($TSC$) and calculate the transmission start time ($TimeStart$) based on a reserved time ($RT$) and the $TSC$. The $TimeStart$ would be calculated as follows: $TimeStart = TSC - TSC \% RT + 2 * RT$. For instance, if $TSC$ obtained by the sender is 39243466, assuming $RT$ is 50000, then $TimeStart$ is 39300000. Thereafter, starting from $TimeStart$, each transmission cycle($T_s$) the sender transmits one data and the receiver receives one data.

\subsubsection{Transmission}
\label{section4.2.2}
After synchronization, every $T_s$ the sender runs different loops to transmit the information according to the data sent and the receiver runs nop-loop to receive the information. For the execution time differences caused by different instructions, we have selected four loops based on the results in Section~\ref{section3.1.2} to create the DI covert channel, which can transmit $2$ bits at a time. The structure of the four loops is shown in Figure~\ref{f4}, and these loops consist of the following instructions: adc, add, xchg, and nop. For the execution time differences generated by the same execution, we have developed the SI covert channel, which can transmit $1$ bit at a time. The loops used by the sender for transmitting information remain the same as described in Section~\ref{section3.2.2}. The transmission protocols for the sender and receiver are outlined in Algorithm~\ref{a1} and Algorithm~\ref{a2}.

\begin{algorithm}[!t]
  \caption{Sender Covert Channel Protocol}
  \label{a1}
  \LinesNumbered
  \KwIn{$len$, $m[n]$, $T_s$, $loop_j$}
  //$len$ is the number of bits sent by the sender at one
  
  //time;

  //$m[n]$ is the $len~*~n$ bits long message to transfer 

  //on the channel;

  //$T_s$ is the transmission period;

  //$loop_j$ is the specific loop, which transfers data $j$;

  //$TimeStart$ is the start time of the whole
  
  //transmission;

  //$TSC$ is current time stamp counter, obtained by
  
  //$rdtsc$;

  $T_{start}$=$TimeStart$;

  \For{$i~=~0;~i~<~n;~i++$}{
    \While{$TSC~<~T_{start}$}{
      nothing;
    }
    \If{$m[i]~==~j$}{
      Execute $loop_j$;
    }
    $T_{start}~=~T_{start}~+~T_s$;
  }
\end{algorithm}

\begin{algorithm}[!t]
  \caption{Receiver Covert Channel Protocol}
  \label{a2}
  \LinesNumbered
  \KwIn{$len$, $T_s$, $ETR_j$}
  //$len$ is the number of bits received by the receiver at
  
  //one time; 

  //$T_s$ is the transmission period;

  //$ETR_j$ is the execution time range of nop-receiver 
  
  //of the receiver when the sender transfers $j$;

  //$TimeStart$ is the start time of the whole
  
  //transmission;

  //$TSC$ is current time stamp counter, obtained by 
  
  //$rdtsc$;

  $T_{start}$ = $TimeStart$;

  \For{$i~=~0; i~<~n;~i++$}{
    \While{$TSC~<~T_{start}$}{
      nothing;
    }

    Execute nop-loop and time the execution;
    \If{$execution\_time~belongs~to~ETR_j$}{
      $res$ = $j$;
    }

    Infer that the transmitted $len$ bits are $res$;

    $T_{start}~=~T_{start}~+~T_s$;
  }
\end{algorithm}

\begin{figure}[!t]
  \centering
  \includegraphics[width=3in]{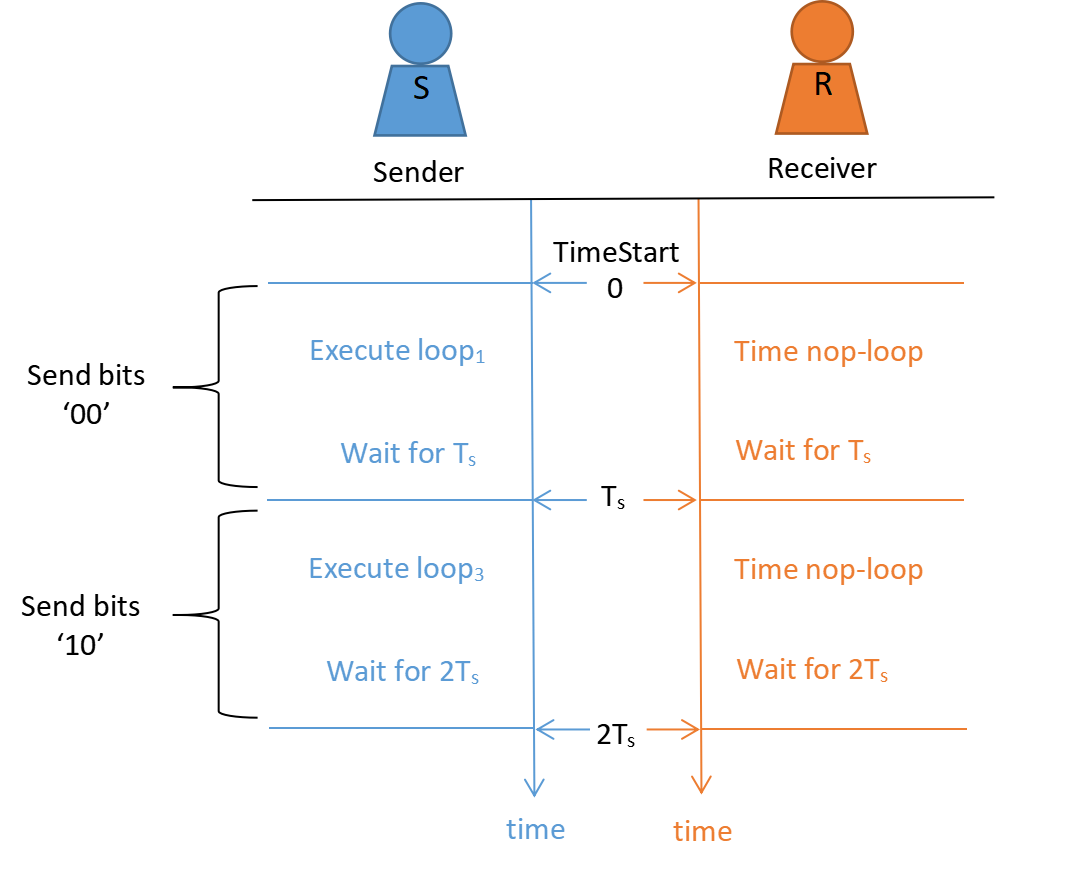}
  \caption{Example of transmission.}
  \label{f10}
\end{figure}

Figure~\ref{f10} illustrates an example of data transfer between the sender and receiver using the covert channels. The process begins with the sender and receiver synchronizing and obtaining the value of $TimeStart$. Once synchronized, the sender runs a specific $loop_j$ every $T_s$ according to the data $j$ to be sent, and the receiver gets nop-loop's execution time $execution\_time$ every $T_s$. Based on $ETR_j$ (the execution time of the receiver's nop-loop under the specific $loop_j$ run by the sender) and the measured $execution\_time$, the receiver decodes the transmitted data $res$.

\subsection{Evaluation}
\label{section4.3}
\subsubsection{Evaluation environment}
\label{section4.3.1}
We next evaluate the performance of the two covert channels introduced in this paper within the specified evaluation environment outlined in Table~\ref{table1}. To implement the covert channels, both the sender and receiver are assigned to two logical cores of the same physical core using the sched\_setaffinity. The execution time of the nop-loop and the TSC are measured by the rdtsc instruction.

\begin{table}[!t]
  \centering
  \caption{Evaluation environment.}
  \label{table1}
  \begin{tabular}{|c|c|}
  \hline
  Processor         & i7-10700       \\ \hline
  Microarchitecture & Cometlake      \\ \hline
  Num of cores      & 16             \\ \hline
  Frequency         & 2.9GHz         \\ \hline
  OS                & Ubuntu 20.04.3 \\ \hline
  \end{tabular}
\end{table}

\subsubsection{Performance evaluation}
\label{section4.3.2}
In this subsection, we assess the performance of both the DI covert channel and SI covert channel. The evaluation metrics used are consistent with previous studies~\cite{12yao2018coherence,25xu2022reverse} and include accuracy and bandwidth. The accuracy represents the ratio of accurately transmitted data to the total amount of transmitted data. The bandwidth metric indicates the transmission capability of the covert channels. It is calculated by dividing the CPU frequency by the $T_s$, i.e. $bandwidth = frequency / T_s$.

(1) Influence of parameters

In the DI covert channel, the parameters that can impact the performance include $T_s$ and the number of iterations of the receiver's loop and sender's loop. Among them, $T_s$ directly influences the bandwidth of the covert channel, and a smaller $T_s$ results in a larger bandwidth. Conversely, a larger $T_s$ reduces the bandwidth but provides greater resilience to noise, as it allows the transmission in one cycle to have minimal impact on the transmission in the subsequent $T_s$. The number of iterations of the receiver's nop-loop affects the accuracy of the covert channel. A larger number of iterations leads to larger time differences in the nop-loop execution, which can enhance accuracy. However, if the number of iterations is too larger, there is a risk of exceeding $T_s$ due to noise, thereby interfering with the transmission in subsequent $T_s$. The number of iterations of the sender's loop should be set such that the execution time of the sender's loop is greater than the execution time of the receiver's nop-loop to ensure accurate transmission. However, a larger number of iterations of the sender's loop also increases the likelihood of exceeding $T_s$ due to noise, which can affect the transmission in the subsequent $T_s$.

Therefore, we configure the iterations of the sender's loop to a value that aligns with the execution time of the receiver's loop, and evaluate the performance of the DI covert channel by varying the value of $T_s$ and the iterations of the receiver's loop.

In the SI covert channel, the performance is also influenced by the parameter $T_s$. However, unlike the DI covert channel, the execution time difference of the receiver's nop-loop is not affected by the iterations of the receiver's loop. Instead, it is influenced by the number of Cache lines utilized by the sender.

Therefore, we set the iterations of the receiver's nop-loop to cover the execution time of the sender's loop and evaluate the performance of the SI covert channel by varying $T_s$ and the number of Cache lines utilized by the sender.

(2) DI covert channel

Because the execution time difference of nop-loop in Section~\ref{section3.2.2} is obvious and the execution time remains within the range of $1,900$ cycles, we set $T_s$ to $3,000-10,000$ cycles and the number of iterations of the receiver's nop-loop to $50$ and $100$.

\begin{figure}[!t]
  \centering
  \includegraphics[width=3in]{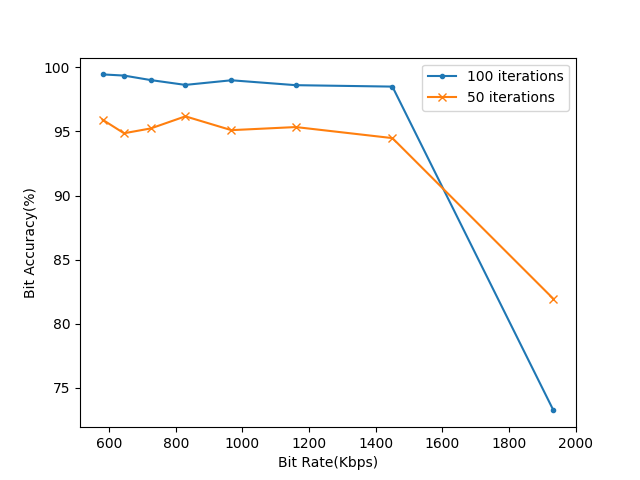}
  \caption{Performance of DI covert channel.}
  \label{f11}
\end{figure}

We tested the accuracy and bandwidth of the DI covert channel by performing $500,000$ transmissions, and the results are shown in Figure~\ref{f11}. The results demonstrate that when $T_s$ is set within the range of $4,000-10,000$ cycles, the accuracy achieved with $100$ iterations of the receiver's loop surpasses that of $50$ iterations. Moreover, as $T_s$ gradually decreases from $10,000$ cycles to $4,000$ cycles, the accuracy shows a slight decline, suggesting that even in the presence of noise, $T_s$ is still sufficient to encompass the execution time of both the receiver's and sender's loops. The primary factor influencing accuracy is the difference in execution time observed in the receiver's nop-loop. However, a significant drop in accuracy occurs when $T_s$ is set to $3,000$ cycles in both cases. This indicates that $T_s$ is no longer able to accommodate the execution time of the receiver's and sender's loops under the influence of noise. Nevertheless, the loop with $50$ iterations experiences less impact from noise as its execution time is shorter, resulting in a higher accuracy compared to the case with $100$ iterations.

(3) SI covert channel

According to Section~\ref{section3.2.2}, the receiver's nop-loop exhibits a noticeable time difference only when the sender utilizes $60$ or more Cache lines. Additionally, considering that the sender needs to flush the Cache in each transmission cycle, resulting in an extended execution time, we set $T_s$ to $6,000-16,000$ cycles and the number of Cache lines utilized to $60$ and $70$ for testing.

\begin{figure}[!t]
  \centering
  \includegraphics[width=3in]{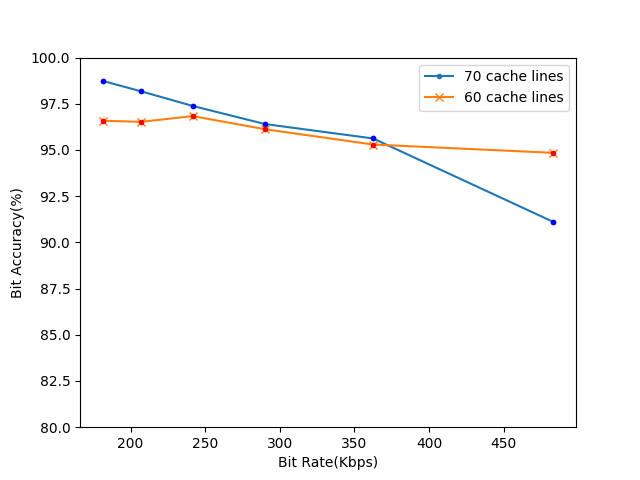}
  \caption{Performance of SI covert channel.}
  \label{f12}
\end{figure}

We test the accuracy and bandwidth of the SI covert channel by performing $500,000$ transmissions, and the results are presented in Figure~\ref{f12}. It can be seen that when $T_s$ is set to $8,000-16,000$ cycles, the accuracy achieved with $70$ Cache lines utilized by the sender is superior to that achieved with $60$ Cache lines utilized. As $T_s$ gradually decreases from $16,000$ to $8,000$ cycles, the accuracy gradually decreases as well. This indicates that even in the presence of noise, $T_s$ is capable of accommodating the execution time of both the receiver's and sender's loops. The primary factor influencing accuracy is the difference in the execution time of the receiver's nop-loop. Notably, when $T_s$ is $6,000$ cycles, a significant drop in accuracy is observed with $70$ Cache lines utilized by the sender, suggesting that $T_s$ is insufficient to encompass the execution time of the receiver's and sender's loops under the influence of noise. However, when the sender utilizes $60$ Cache lines, $T_s$ remains effective in encompassing the execution time of both the receiver's and sender's loops, resulting in higher accuracy.

\subsubsection{Overall evaluation}
\label{section4.3.4}
The overall performance of the two covert channels proposed in this paper is summarized in Table~\ref{table2}. Based on the evaluation results presented in Section~\ref{section4.3.2}, it is observed that the DI covert channel outperforms the SI covert channel in the performance. The DI covert channel achieves a maximum bandwidth of $1450$ Kbps with an accuracy of $98.5\%$. When compared to previous covert channels created on Intel processors~\cite{3taram2022secsmt,10gruss2016flush+,11xiong2020leaking,12yao2018coherence,15guo2022adversarial,19schwarz2019netspectre,23ren2021see,24deng2022leaky,25xu2022reverse,26aldaya2019port,29wang2023bandwidthbreach,42paccagnella2021lord}, the DI covert channel exhibits higher bandwidth than most of the existing covert channels, except for Cache, Prefetcher, Line Fill Buffer, Fetch Bandwidth, and Ring.

However, various protections are proposed against existing covert channels, such as partitioning~\cite{3taram2022secsmt}, our proposed channels can bypass these protections by leveraging retirement. 

\begin{table}[!t]
  \centering
  \caption{Overall performance of two covert channels.}
  \label{table2}
  \begin{tabular}{|c|c|c|}
  \hline
  Covert channel                                                                                                           & DI                                                          & SI                                                            \\ \hline
  Best accuracy                                                                                                            & 99.46\%                                                     & 95.92\%                                                       \\ \hline
  Bandwidth at best accuracy                                                                                               & 580Kbps                                                     & 181.25 Kbps                                                   \\ \hline
  \begin{tabular}[c]{@{}c@{}}Maximum bandwidth and\\corresponding accuracy\\when accuracy \textgreater{}90\%\end{tabular} & \begin{tabular}[c]{@{}c@{}}1450 Kbps \\ 98.5\%\end{tabular} & \begin{tabular}[c]{@{}c@{}}483.33 Kbps\\ 94.85\%\end{tabular} \\ \hline
  \end{tabular}
\end{table}

\subsubsection{Comparison with normal execution}
\label{section4.3.3}

The detection method utilizing HPC relies on the discrepancies in the usage of microarchitectural resources between the normal execution and the covert channels to detect the covert channels. The larger the difference in resource utilization between the normal execution and the covert channels, the more readily detectable the covert channel becomes through HPC-based detection method.

For the two new covert channels proposed in this paper, the sender of the DI covert channel is a loop with varying retirement usage, which is a partial truncation of the normal execution and therefore similar to the normal execution. The sender of the SI covert channel have a large number of Cache miss during transmission, resulting in noticeable differences compared to normal execution. The receivers of DI covert channels and SI covert channels constantly utilize the retirement, leading to possible differences in retirement usage compared to normal program execution. These differences in retirement usage can be detected using HPC-based detection method. Hence, we evaluate the differences in this subsection.

\begin{figure}[!t]
  \centering
  \includegraphics[width=3in]{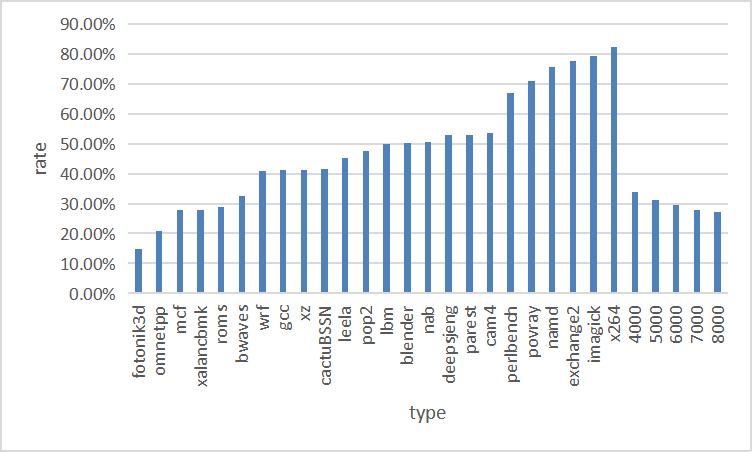}
  \caption{Retirement usage of different programs.}
  \label{f13}
\end{figure}

We choose the benchmarks from the SPEC CPU$2017$ suite as examples of normal execution and nop-loop with $100$ iterations and $T_s$ from $4,000$ to $8,000$ as covert channel examples. The results are shown in Figure~\ref{f13}.

The results indicate that the retirement usage of the receiver is relatively low and similar to certain benchmarks. This can be attributed to the instruction that waits for the start of the next transmission cycle. As a result, the DI covert channel is less likely to be detected by the HPC-based detection methods. However, the sender of the SI covert channel is easily detected by the HPC-based detection methods.

\section{Other applications of retirement leakage}
\label{section5}

In this section, two logical cores on the same physical core are referred to as the victim and the attacker to explore further applications of the retirement leakage. The applications in this section are implemented on the Cometlake as shown in Table~\ref{table1}.

\subsection{Spectre attack variant}
\label{section5.1}
According to~\cite{2agner2021microarchitecture}, the branch misprediction penalty varies a lot in the Cometlake. We speculate that this penalty is related to the number and type of instructions speculated to be executed in the ROB. In light of this, we have designed two different loops loop1 (decoding out $1$ jump instruction per cycle) and loop2 (decoding out $1$ jump instruction and $1$ add instruction per cycle) to investigate the penalty variations. Through testing, we have observed that incorrectly speculating the execution of loop1 generates a penalty time of $68$ cycles and incorrectly speculating the execution of loop2 generates a penalty cycle of $57$ cycles. Exploiting this behavior, the victim can execute either loop1 or loop2 in the speculative execution phase depending on the secret obtained, which eventually generates different penalties in the fallback. During the fallback, the retirement is in a stalling state. Therefore, the attacker can infer the victim's fallback penalty period by testing the available time of its own retirement, and thus deduce the secret leaked by the victim.

\begin{figure}[!t]
  \centering
  \includegraphics[width=3in]{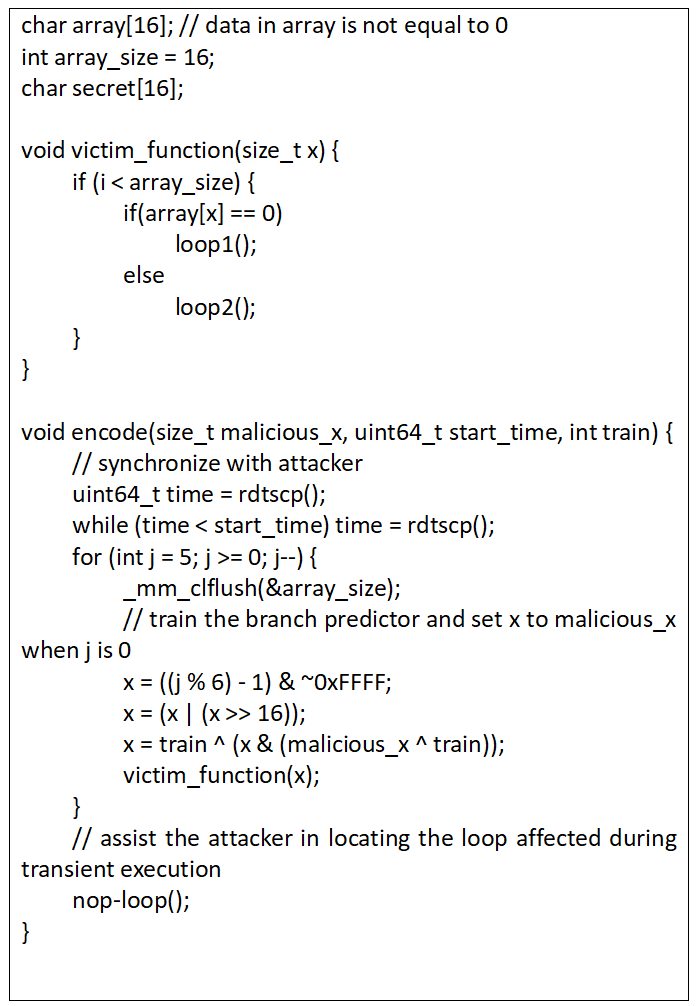}
  \caption{Victim code snippet in the new variant of Spectre v1.}
  \label{f14}
\end{figure}

\begin{figure}[!t]
  \centering
  \includegraphics[width=3in]{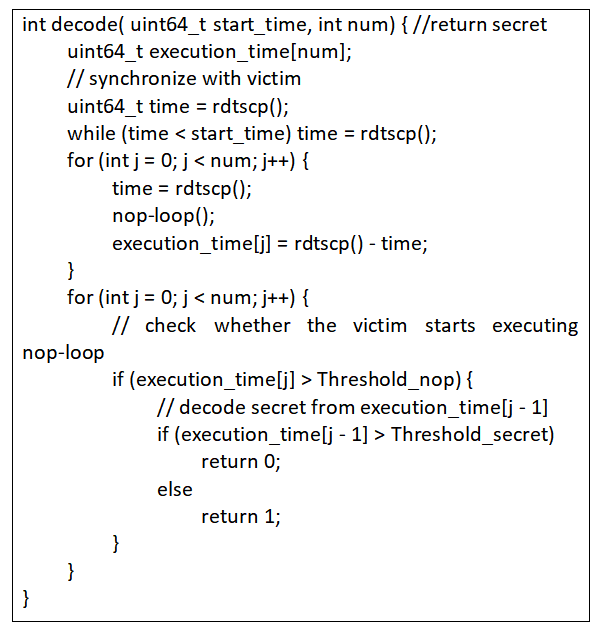}
  \caption{Attacker code snippet in the new variant of Spectre v1.}
  \label{f15}
\end{figure}

Based on the above principles, we apply the retirement leakage to the Spectre attacks and propose a new variant of Spectre v1. The code snippets of the victim and attacker are illustrated in Figure~\ref{f14} and Figure~\ref{f15}. Similar to Spectre v1, the victim needs to train the branch predictor to induce a branch misprediction in the $victim\_function$ and flush the $array\_size$ to prolong the transient execution window. However, in contrast to Spectre v1, our variant requires the victim and attacker to synchronize at the beginning of each attack, which can be achieved through the timestamps. When $j$ equals $0$, the value of $x$ is changed to $malicious\_x$. The victim bypasses the boundary detection by misprediction and leverages $array[x]$ to access the private information. Furthermore, the victim executes different loops during transient execution based on the leaked private information to induce varying degrees of fallback time. Meanwhile, the attacker continuously collects the execution time of the nop-loop, so it needs to identify the nop-loop that reflects the fallback time of the victim. We locate the nop-loop by having the victim execute a nop-loop after the attack is completed (the same can be done with the other instructions that have different retirement usage). After that, the secret can be decoded according to the execution time of this nop-loop, i.e., the private information is not $0$ if the time is shorter than a certain threshold, otherwise the private information is $0$. Based on the above transmission method, the retirement leakage can be extended to other transient attacks.

We set our Spectre attack to $20,000$ cycles per attack, and the attacker runs nop-loop $20$ times in one attack, i.e., $num$ is $20$, and the iterations of the nop-loop are $100$. To mitigate the influence of noise and the correct prediction of the branch predictor, we aggregate the results every $5$ attacks, improving the overall attack accuracy. The final attack result is determined by voting. 
Finally, we conducted an evaluation of our Spectre attack by performing $200,000$ attacks. The results show that our Spectre attack achieves a bandwidth of $29$ Kbps with an accuracy rate of $94.17\%$. Compared to existing Spectre attacks that give specific performance~\cite{23ren2021see,25xu2022reverse}, the accuracy of our attack is inferior to some existing attacks, but it compensates by offering a better bandwidth than these counterparts.

\begin{table}[!t]
  \centering
  \caption{L1D miss rate of Spectre v1 attacks.}
  \label{table3}
  \begin{tabular}{|c|ccc|cc|}
  \hline
                                                                & \multicolumn{3}{c|}{Existing Work}                                                                                                                                            & \multicolumn{2}{c|}{Our Work}          \\ \hline
                                                                & \multicolumn{1}{c|}{\begin{tabular}[c]{@{}c@{}}F+R\\ MEM~\cite{11xiong2020leaking}\end{tabular}} & \multicolumn{1}{c|}{\begin{tabular}[c]{@{}c@{}}F+R\\ L1D~\cite{11xiong2020leaking}\end{tabular}} & LRU~\cite{11xiong2020leaking} & \multicolumn{1}{c|}{Victim} & Attacker \\ \hline
  \begin{tabular}[c]{@{}c@{}}L1D Miss\\Rate\end{tabular} & \multicolumn{1}{c|}{2.86\%}                                                    & \multicolumn{1}{c|}{4.84\%}                                                    & 4.13\%      & \multicolumn{1}{c|}{0.29\%} & 0.04\%   \\ \hline
  \end{tabular}
  \end{table}

Existing Cache-based Spectre v1 attacks predominantly employ L1D to encode sensitive information. However, these attacks tend to cause a substantial number of L1D misses, which can make them susceptible to detection using HPC-based methods. We conducted tests to evaluate the L1D miss rate in our attack, and the findings are summarized in Table~\ref{table3}. To provide a comprehensive comparison, we also included L1D miss rate data from existing Cache-based Spectre v1 attacks (derived from~\cite{11xiong2020leaking}). The results indicate that our attack exhibits a relatively low L1D miss rate in comparison to existing attacks. This implies that our attack is less likely to be detected through L1D miss rate analysis. Furthermore, our attack leverages the retirement transmission information to bypass the protection against existing covert channels.

\subsection{Inferring user's programs}
\label{section5.2}
Distinct programs encompass unique instruction sequences, with variations in the execution cycles of individual instructions within these sequences. Consequently, the retirement experiences varying degrees of stalling depending on the specific instruction being executed. Exploiting this disparity, an attacker can assess the stalling situation of the victim's retirement based on the availability of its own retirement, thereby inferring the programs being executed by the victim.

\begin{figure*}[!t]
  \centering
  \subfloat[perlbench]{\includegraphics[width=2in]{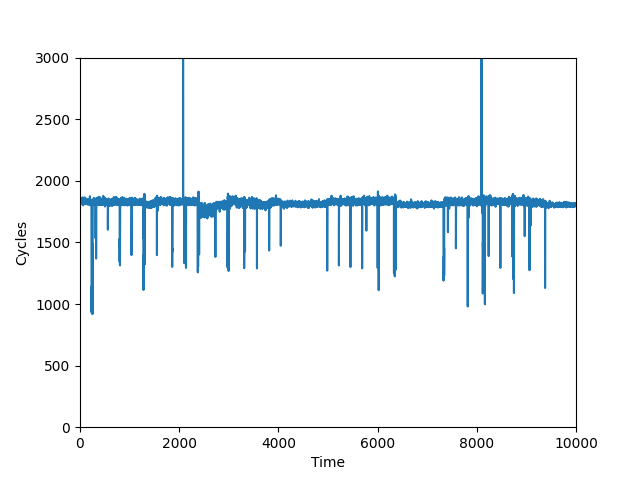}
  \label{fig16a}}
  \hfil
  \subfloat[gcc]{\includegraphics[width=2in]{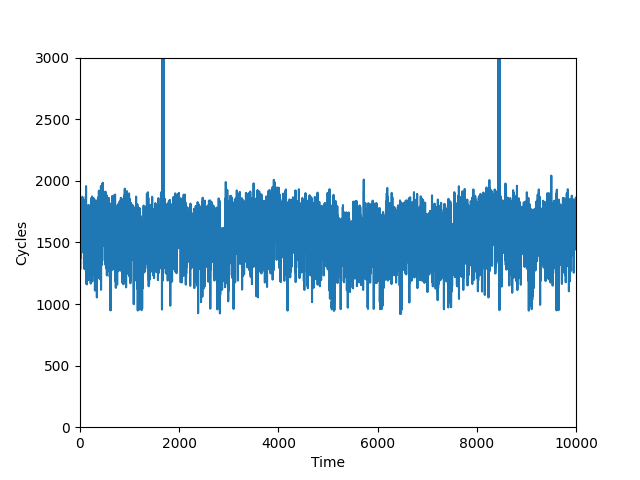}
  \label{fig16b}}
  \hfil
  \subfloat[omnetpp]{\includegraphics[width=2in]{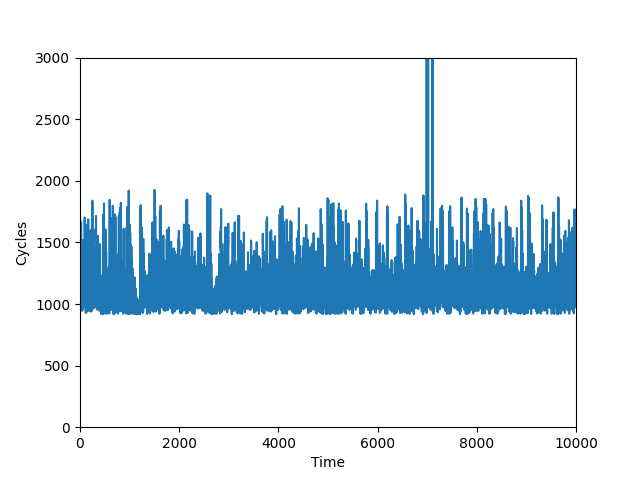}
  \label{fig16c}}
  \caption{Execution time of attacker's nop-loop over time when running different programs on the victim.}
  \label{fig16}
\end{figure*}

In this paper, we conduct experiments using 10 distinct integer benchmarks from SPEC CPU$2017$ as the target programs. To assess the attacker's own retirement usage, we employ the nop-loop as the receiver code. Figure~\ref{fig16} shows the variation of the execution time of the nop-loop on the attacker with sampling time when the victim runs three different integer benchmarks. The results clearly demonstrate significant differences in the execution time of the nop-loop on the attacker when the victim executes different programs. 

Based on the above principle, we conduct an extensive experimentation process to investigate the feasibility of inferring different programs using retirement. We take every $2,000$ nop-loop execution times as a sample and collect 6313 traces corresponding to $10$ different benchmarks, and use CNN to model the sampling results corresponding to different integer benchmarks to distinguish different programs. Because the instruction types in different programs may change significantly, resulting in some benchmarks containing multiple different retirement usage, we use only one of the retirement usage cases as a representative to verify that retirement units can be used to infer different programs. In addition, due to the presence of similar retirement usage patterns among certain benchmarks, the inference success rate is negatively affected. So we also test the inference success rate when some different benchmarks are grouped into one category. We finally test the inference success rate when the number of benchmarks running on the victim varied, and the results are shown in Figure~\ref{f17}. The specific benchmarks employed on the victim can be found in Appendix~\ref{AppendixA}, and the preprocessing methodology, the ratio of training set to test set division, and the detailed structure of CNN are provided in Appendix~\ref{AppendixB}.

\begin{figure}[!t]
  \centering
  \includegraphics[width=3in]{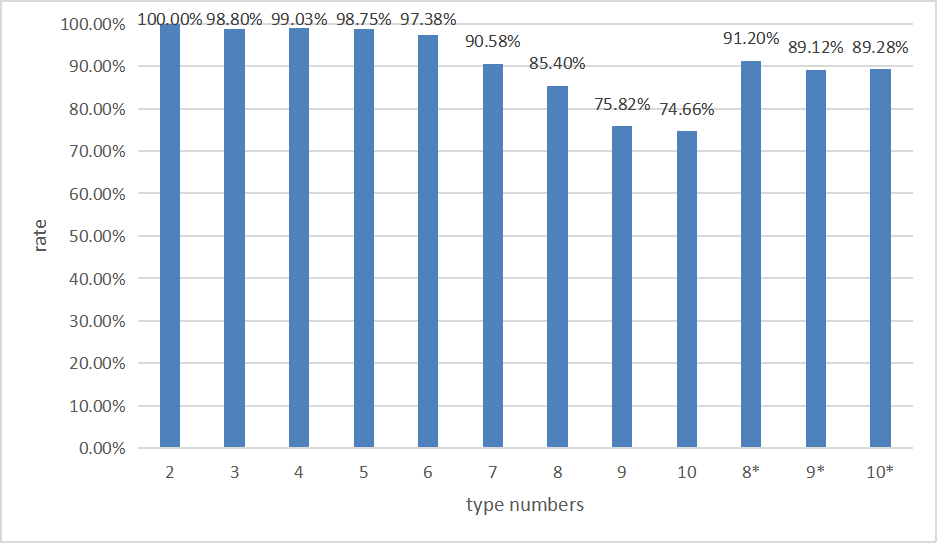}
  \caption{Accuracy rate when the number of inferred programs varies. * represents some of the benchmarks are grouped together.}
  \label{f17}
\end{figure}

The results reveals that when the victim runs $6$ different benchmarks and below, the attacker can successfully infer the programs run by the victim with a high accuracy. However, as the number of benchmarks executed by the victim exceeds $7$, the accuracy of the attacker's inference drastically declines. This decline in accuracy can be attributed to the similarity in retirement usage patterns among some benchmarks. Without the ability to align traces, the inference process becomes unreliable. However, when we classify some benchmarks with similar characteristics into one category, the final inferred accuracy of the $10$ benchmarks can still reach $89.28\%$.

\section{Discussion of possible mitigations}
\label{section6}

Among the existing mitigations, victims can use partitioning and detection using HPC against the novel covert channels introduced in this paper.

(1) Partitioning

One effective approach to prevent the covert channels proposed in this paper is partitioning the retirement by time. Among the various strategies available, static partitioning is the most fundamental one, letting two logical cores take turns using retirement. By adopting this strategy, it becomes infeasible for the sender to influence the receiver through retirement. However, it comes with the drawback of underutilizing retirement and reducing overall processor performance. To mitigate this performance impact, dynamic partitioning methods can be implemented, such as contention-driven partitioning~\cite{4townley2019smt} and asymmetric SMT~\cite{3taram2022secsmt}.

(2) Detection using HPC

Section~\ref{section4.3.3} reveals that when the detection granularity is set to one transmission cycle, the victim is unable to detect the presence of the DI covert channel through retirement usage. However, as indicated by the results presented in Section~\ref{section3.1.2}, the retirement usage of the receiver's nop-loop in the DI covert channel is close to $100\%$, which deviates from normal execution patterns. Consequently, if the detection granularity is increased to $1,000$ cycles, which is shorter than the execution time of the nop-loop, the receiver of the DI covert channel can theoretically be detected using HPC, but there will be a false alarm problem. Moreover, the policy will visit the HPC frequently (once in $1,000$ cycles), which can negatively impact processor performance.

\section{Conclusion}
\label{section7}

In this paper, we found that there is a security vulnerability in the retirement in Intel processors and created two new covert channels that effectively bypass existing protection policies. Among them, the DI covert channel can achieve $98.5\%$ accuracy with a bandwidth of $1450$ Kbps, and the SI covert channel can achieve $94.85\%$ accuracy with a bandwidth of $483.33$ Kbps. Furthermore, we have extended the application of the retirement to Spectre attacks to propose a novel variant of Spectre v1, and propose an attack method that uses retirement to infer the programs run by the victim. Finally, we discuss possible mitigations against our covert channels.

In the future, we will further investigate the potential applications of retirement in detection. In some specific devices, such as industrial control devices, the processor primarily executes designated programs, but an attacker may inject malicious programs into these devices to carry out the attack. The execution time of the nop-loop, as described in Section~\ref{section5.2}, can be used to model the behavior of authorized programs on these devices. By establishing these models, it becomes possible to detect the presence of injected malicious programs by testing the actual programs running on the devices.

\section*{Acknowledgements}
This work is supported by National Key R\&D Program of China, No. 2022YFB3103800 and
National Natural Science Foundation of China (NSFC), Nos. 61972295 and 62072247.

\bibliographystyle{IEEEtranS}
\bibliography{new}

\appendices
\section{}
\label{AppendixA}
In Section~\ref{section5.2}, the benchmarks represented by the number 
of different benchmark categories are shown below:

$2$: perlbench and gcc.

$3$: perlbench, gcc and mcf.

$4$: perlbench, gcc, mcf and xz.

$5$: perlbench, gcc, mcf, xz and xalancbmk.

$6$: perlbench, gcc, mcf, xz, xalancbmk and omnetpp.

$7$: perlbench, gcc, mcf, xz, xalancbmk, omnetpp and exchange2.

$8$: perlbench, gcc, mcf, xz, xalancbmk, omnetpp, exchange2 and x264.

$9$: perlbench, gcc, mcf, xz, xalancbmk, omnetpp, exchange2, x264 and leela.

$10$: perlbench, gcc, mcf, xz, xalancbmk, omnetpp, exchange2, x264, leela and deepsjeng.

$8*$ is to group perlbench and x264 together, $9*$ is to 
group xalancbmk and leela together on the basis of $8*$, 
10* is to group xalancbmk, leela and deepsjeng together 
on the basis of $8*$.

\section{}
\label{AppendixB}

\begin{table}[!t]
  \centering
  \caption{Structure of CNN.}
  \label{table4}
  \begin{tabular}{|c|c|c|c|}
  \hline
  Layer   & Input     & Parameters                                                                            & Output    \\ \hline
  Reshape & (2000, 1) & Reshape(10)                                                                           & (200, 10) \\ \hline
  Conv\_1 & (200, 10) & \begin{tabular}[c]{@{}c@{}}Conv(64, 5, leaky\_relu)\\ MaxPooling(2)\\ BN\end{tabular} & (98, 64)  \\ \hline
  Conv\_2 & (98, 64)  & \begin{tabular}[c]{@{}c@{}}Conv(64, 5, leaky\_relu)\\ MaxPooling(2)\\ BN\end{tabular} & (47, 64)  \\ \hline
  GRU     & (47, 64)  & \begin{tabular}[c]{@{}c@{}}GRU(128) \\ BN\end{tabular}                                & (47, 128) \\ \hline
  FC      & (47, 128) & FC                                                                                    & (6016)    \\ \hline
  Output  & (6016)    & Dense(c, softmax)                                                                     & (c)       \\ \hline
  \end{tabular}
  \end{table}

To preprocess the data, we performed a subtraction of 
the average value of the sampled data before data 
training. Out of the total $6313$ traces, $70\%$ of them 
were allocated as the training set for the purpose of 
training, while the remaining $30\%$ were reserved for 
accuracy testing. The architecture of CNN we utilized is 
outlined in Table~\ref{table4}. It consists of two convolutional 
layers, a Gated Recurrent Unit (GRU) layer, and a 
flatten layer. Where, $c$ is the final number of 
classifications.

\end{document}